\documentclass[letterpaper]{article} 
\usepackage{aaai25}  
\usepackage{times}  
\usepackage{helvet}  
\usepackage{courier}  
\usepackage[hyphens]{url}  
\usepackage{graphicx} 
\usepackage{natbib}  
\usepackage{caption} 
\usepackage{algorithm}
\usepackage{algorithmic}

\usepackage{newfloat}
\usepackage{listings}
\DeclareCaptionStyle{ruled}{labelfont=normalfont,labelsep=colon,strut=off} 
\floatstyle{ruled}
\newfloat{listing}{tb}{lst}{}
\floatname{listing}{Listing}
\title{EduBot – Can LLMs Solve Personalized Learning and Programming Assignments?}

\title{EduBot – Can LLMs Solve Personalized Learning and Programming Assignments?}
\author {
    Yibin Wang\equalcontrib\textsuperscript{\rm 1,},
    Jiaxi Xie\equalcontrib\textsuperscript{\rm 1},
    Lakshminarayanan Subramanian\textsuperscript{\rm 1}
}
\affiliations {
    \textsuperscript{\rm 1}Courant Institute of Mathematical Sciences, New York University\\
    yw4145@nyu.edu, jx1210@nyu.edu, lakshmi@cs.nyu.edu
}
\usepackage{bibentry}

\begin{document}

\maketitle

\begin{abstract}
The prevalence of Large Language Models (LLMs) is revolutionizing the process of writing code. General and code LLMs have shown impressive performance in generating standalone functions and code-completion tasks with one-shot queries. However, the ability to solve comprehensive programming tasks with recursive requests and bug fixes remains questionable. In this paper, we propose EduBot, an intelligent automated assistant system that combines conceptual knowledge teaching, end-to-end code development, personalized programming through recursive prompt-driven methods, and debugging with limited human interventions powered by LLMs. We show that EduBot can solve complicated programming tasks consisting of sub-tasks with increasing difficulties ranging from conceptual to coding questions by recursive automatic prompt-driven systems without finetuning on LLMs themselves. To further evaluate EduBot's performance, we design and conduct a benchmark suite consisting of 20 scenarios in algorithms, machine learning, and real-world problems. The result shows that EduBot can complete most scenarios in less than 20 minutes. Based on the benchmark suites, we perform a comparative study to take different LLMs as the backbone and to verify EduBot's compatibility and robustness across LLMs with varying capabilities. We believe that EduBot is an exploratory approach to explore the potential of pre-trained LLMs in multi-step reasoning and code generation for solving personalized assignments with knowledge learning and code generation. 
\end{abstract}

\begin{figure*}[htbp] 
   \centering
   \includegraphics[width=0.7\textwidth]{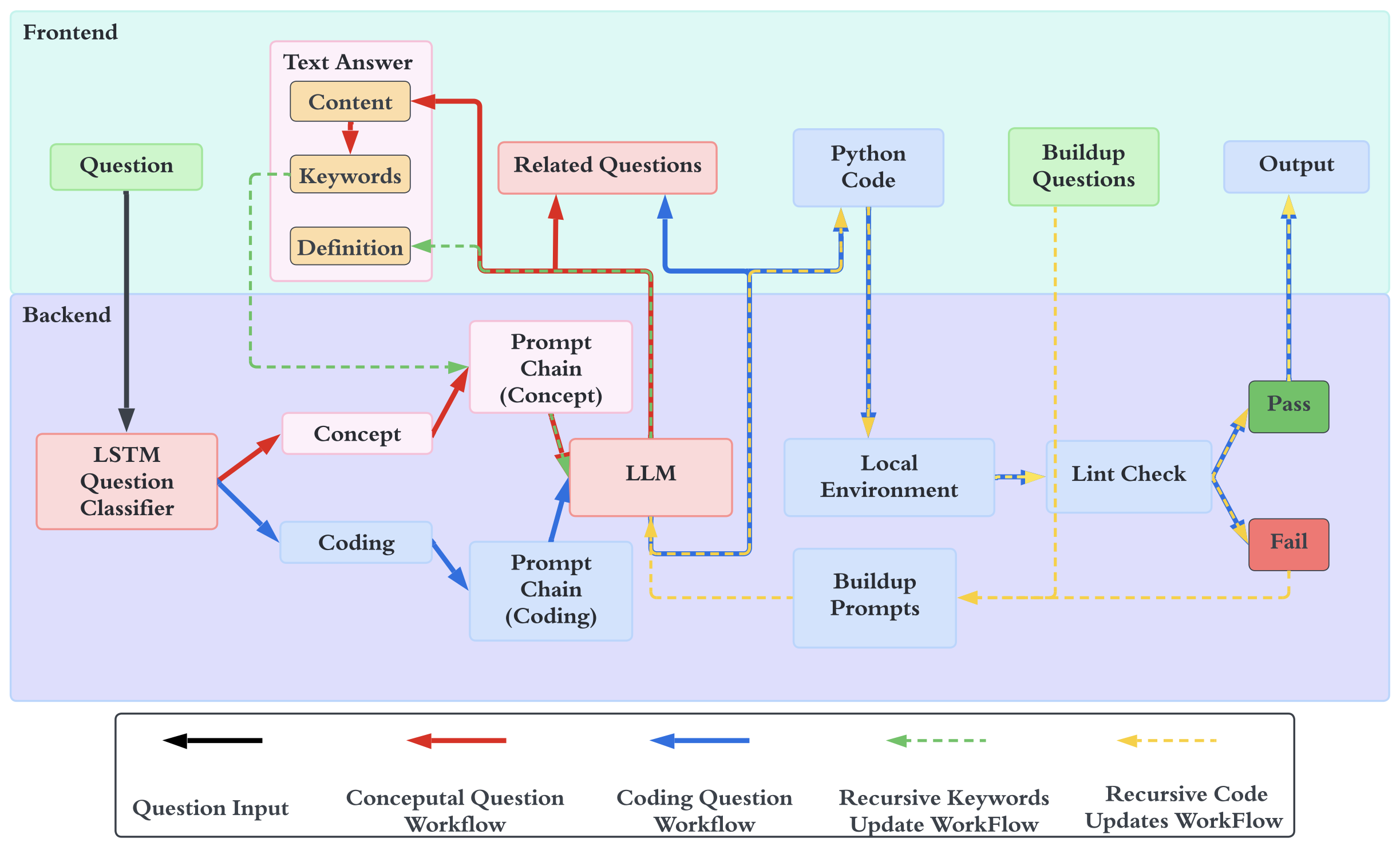} 
   \caption{Architecture and Workflow of EduBot}
   \label{workflow}
\end{figure*}
\section{Introduction}

Recent innovations and advances in Large Language Models (LLMs) such as ChatGPT \cite{OpenAI2022} and Claude \cite{Claude2024} have shown an impressive ability to understand questions in natural languages and provide knowledgeable answers \cite{memarian2023chatgpt, mann2020language}. This prevalence introduces a new approach to solving coding challenges and providing educational assistance through machines, a long-standing open challenge that has attracted a series of researches. The combination of Machine Learning (ML) and Programming by Examples (BPEs) demonstrates the potential of ML models to address this problem \cite{le2014flashextract,gulwani2017programming}. Training or fine-tuning with code corpora have also depicted promising performance on simple one-shot query coding questions and code completion tasks \cite{chen2021evaluating, nijkamp2023codegen,chowdhery2023palm}. Comprehensive calibration metrics for code LLMs on simple tasks, such as code completion, provide practical and in-depth insights for systematic comparisons of different LLMs.

Similarly to previous work to bring online search to code environment to boost programmers' productivity \cite{brandt2010example}, recent work has begun utilizing LLMs to optimize coding workflows, focusing on ensuring syntactic correctness \cite{ugare2024improving}, code understanding \cite{nam2024using}, and test-driven development \cite{fakhoury2024llm}. These efforts aim to address coding problems in a zero-shot manner, without requiring recursive corrections or human interaction. However, LLMs are not perfect and can make mistakes. In many real-world coding scenarios, users might recursively ask the model to improve the code, answering their questions about novel concepts, adding new features, and debugging existing code. Thus, zero-shot question-asking and assistance are not sufficient to handle this situation. 

In this paper, we propose EduBot, an intelligent system powered by LLMs that focuses on conceptual knowledge teaching, recursive prompt-driven programming, and automated debugging to tackle complex tasks comprising subtasks of incremental difficulty. To evaluate EduBot's performance with different LLM backbones, we developed a benchmark suite containing 20 sample scenarios with 79 elementary and advanced sub-tasks. In each scenario, EduBot begins with conceptual questions and progresses to writing complete code based on those concepts, addressing real-world use cases through personalized follow-up queries and recursive debugging. Upon completing a scenario, users gain both a deeper conceptual understanding of specific topics and hands-on coding experience. Additionally, we conducted a comparative study to evaluate the performance of EduBot across various LLM backbones.
\noindent The contributions of this work are as follows: 
\begin{enumerate}
    \item We make the exploratory approach for solving personalized programming tasks with knowledge learning chained by varied levels of sub-tasks with LLMs.
    \item We show that, with limited human intervention and prompting engineering, EduBot can finish most of the complex scenarios from scratch within 20 minutes. 
\end{enumerate}

\section{Related Work}
\subsection{Recursive and Code Generations Beyond the Scope of Simple Code Generation}
To address coding problems with few-shot reasoning, and in alignment with divide-and-conquer principles, the Large Language Model Debugger (LDB) splits coding problems into intermediate states and queries LLMs to write and test each block for code generation \cite{zhong2024ldb}. Moving beyond independent function completions and generations, TOOLGEN \cite{wang2024teaching} focuses on repository-level completion tasks, demonstrating strong performance in resolving dependency coverage issues. By analyzing pooled gradients of outputs with respect to input prompts, North et al. \cite{north2024code} propose a method to trace overlooked requirements and prompt LLMs to address the missing parts of the problem. This enables a recursive workflow and achieves reasonable performance on simple code generation tasks.
\subsection{Concept Knowledge Learning Generation from Online Resources}
CollectiveTeach \cite{ranawat2021collectiveteach} auto-generates lesson plans for computer science courses by leveraging a corpus of approximately 100,000 web pages, highlighting the importance of retrieving valuable online information. Additionally, data mining techniques have been applied to enhance learning by analyzing textbooks and assessing their cognitive load on readers \cite{agrawal2012empowering}.
Quizlet, a flashcard-based study app, offers an effective way to learn fundamental concepts across various subjects. Users can create personalized study sets to review and reinforce their knowledge. Stack Overflow, a popular discussion forum, facilitates knowledge sharing among experienced developers by providing valuable solutions and tips for computer science challenges, spanning both conceptual and coding domains. Similarly, LeetCode offers a collection of algorithmic programming problems, enabling users to practice coding while receiving performance evaluations.

\section{Proposed Solution}
\subsection{Problem Formulation and Workflow}

Learning sorting algorithms such as Bubble Sort, Merge Sort, and Quick Sort involves understanding their mechanisms and analyzing their time and space complexities. This process forms a foundational component of computer science education and can be broken down into meaningful sub-tasks aligned with human learning objectives.

We represent this learning journey as a sample scenario, $Q$, with the optimal solution denoted as $A^{*}(Q)$. Humans naturally decompose such tasks into incremental sub-tasks, which can be represented as a query trace $[q_1, q_2, ..., q_n]$, with corresponding solutions $[a^{*}_1(q_1), a^{*}_2(q_2), ..., a^{*}_n(q_n)]$. For example, the task of learning sorting algorithms can be divided as follows:

\begin{enumerate}
    \item Understand the concept of sorting and its importance.
    \item Learn the mechanisms of common sorting algorithms like Bubble Sort, Merge Sort, and Quick Sort.
    \item Compare these algorithms in terms of time and space complexities.
    \item Implement the algorithms in Python to reinforce understanding.
\end{enumerate}

This task decomposition reflects the human learning process by building on foundational concepts and advancing toward practical implementation. It establishes a structured framework that enables recursive problem-solving systems like EduBot to effectively guide users through the learning process. For more examples for each category, please refer to the Appendix. During actual interactions, users can explore further by asking additional questions to deepen their understanding, extend concepts, or refine code implementations, allowing EduBot to support a personalized and iterative learning journey.

From Figure ~\ref{workflow}, we denote question classifier as $LSTM()$ which takes initial sub-task $q_1$ and output binary values $0$ for conceptual questions and $1$ for coding questions. 

\subsubsection{Concept Questions Workflow}
If $LSTM(q_{i})=0$ where $q_{i}$ is a sub-task of scenario $Q$ and $1 \leq i \leq n$, the respective concept prompt chain, denoted as $Concept()$, which take the sub-task $q_{i}$, will be activated. The output of $Concept(q_i)$ is denoted as $a_{i}(q_{i}) = (c, r)$, where $c$ is denoted as answer of $q_{i}$ and $r$ is denoted as several related questions and answers for practices and further information. From $c$, EduBot extracts a set of keywords $K = \{k_1, k_2, ..., k_n\}$ such that $K \subseteq c$. If a user seeks the explanation for a specific keyword $k_i \in K$, the corresponding definition $d_i$ will be generated. This iterative process refines $a_i(q_i)$ into the optimal answer $a^*_i(q_i)$. 

\subsubsection{Coding Questions Workflow}
If $LSTM(q_i)=1$, then the coding prompt chain $Code()$ will be activated. The output of $Code(q_i)$ is denoted as $a_i(q_i) = (c_{init}, r)$, where $c_{init}$ is the Python code generated by LLM, and $r$ consists of related questions and answers for similar purpose in the concept question workflow. To verify the correctness of $c_{init}$, it is passed to a development environment for a linter check, denoted as $lint()$, to detect syntax errors. The output of $lint(c_{init})$ is either ``pass'' or ``fail''. If the output is ``fail'', a set of error/warning messages, denoted as $M = \{m_1, m_2, ..., m_n\}$, will be displayed in the user interface. 

\subsubsection{Recursive Prompt-Driven Programming and Debugging with Human Inference}
\begin{figure}[htbp]
\includegraphics[width=0.45\textwidth]{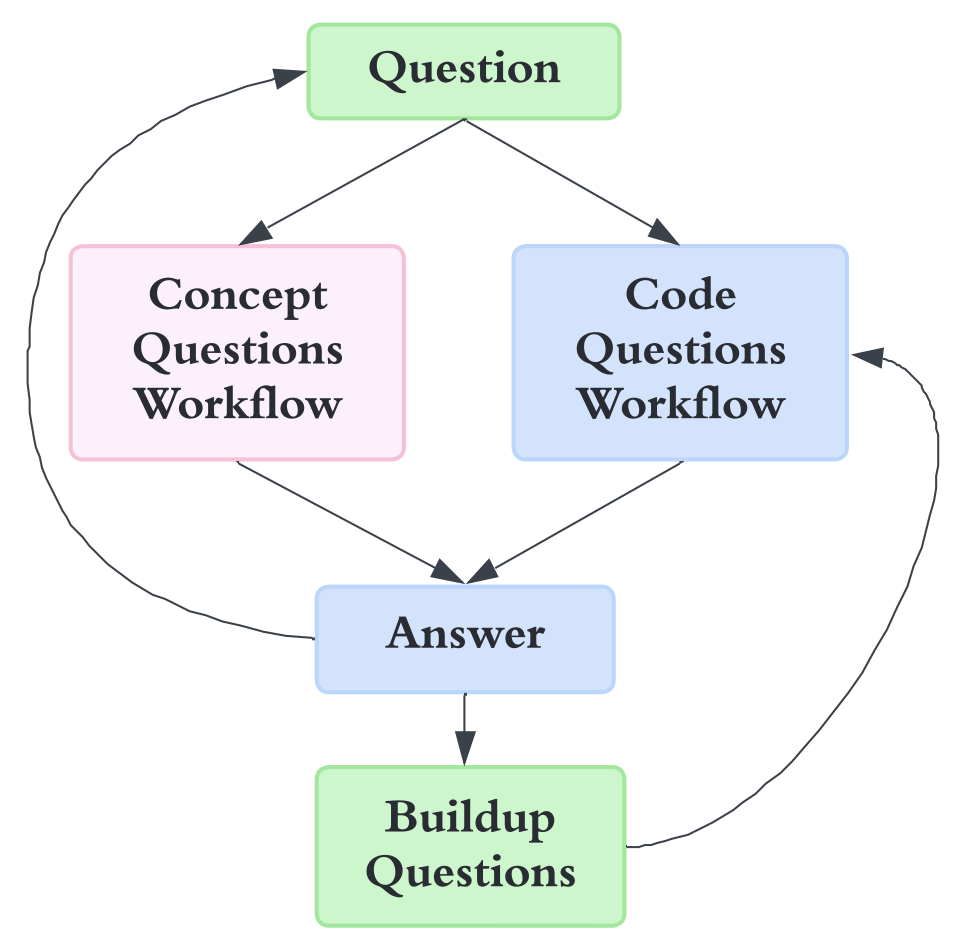}
\caption{Recursive answer improvements}
\label{recursive_worlflow}
\end{figure}
Although current LLMs are capable of tackling various challenging problems, they can still make mistakes. Therefore, it is crucial to enable EduBot to recursively revise and improve its answers generated from either the concept question workflow or the coding question workflow, guided by human input.

In Figure \ref{recursive_worlflow}, EduBot provides buildup prompting system, denoted as $Buildup()$, to handle potential bug fix or personalized feature add-on requests. This system operates based on the answers generated from the coding question workflow or by directly addressing new questions $q$ in the concept question workflow. The buildup prompting systems takes further questions denoted as $q_{buildup}$ and the current code $c_{curr}$ as shown in Equation \ref{eq:buildup}. 
\begin{equation}
\label{eq:buildup}
c_{curr} = Buildup(q_{buildup}, c_{curr})
\end{equation}

\begin{figure}[htbp]
\includegraphics[width=0.45\textwidth]{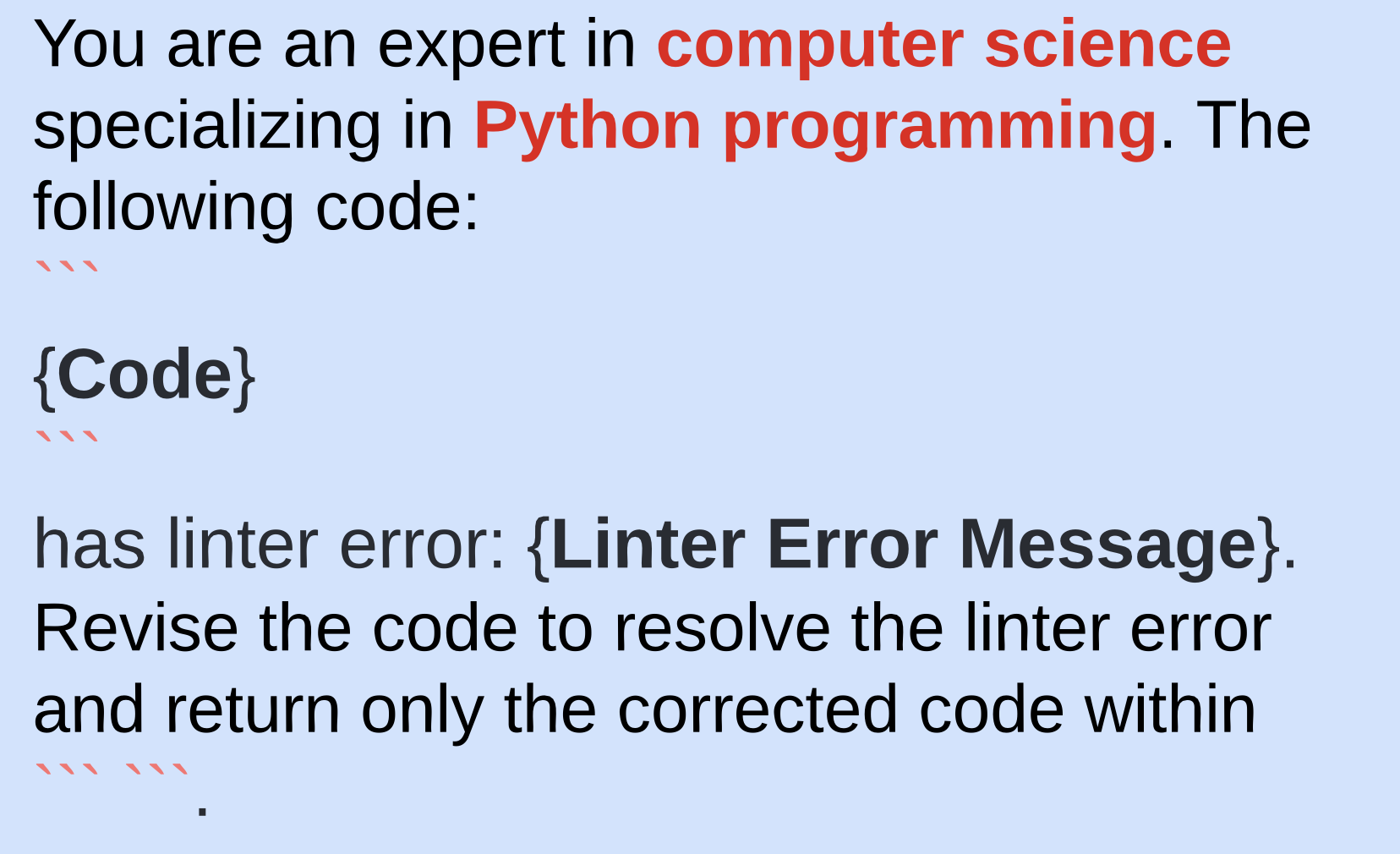}
\caption{Linter error revision prompts}
\label{linter_prompt}
\end{figure}
For the generation of buildup questions in prompting system, if the output of $lint(c_{curr})$ is ``fail'', the user can click on the specific error message $m_i$, where $1 \leq i \leq n$, to activate buildup prompting system. The system will take $m_i$ to automatically construct buildup questions $q_{buildup}$ with current Python code $c_{curr}$ and guide LLM to reassign $c_{curr}$ as Equation \ref{eq:buildup} and Figure \ref{linter_prompt}. This process can be performed recursively until the code passes the lint check. 

\begin{figure}[htbp]
\includegraphics[width=0.45\textwidth]{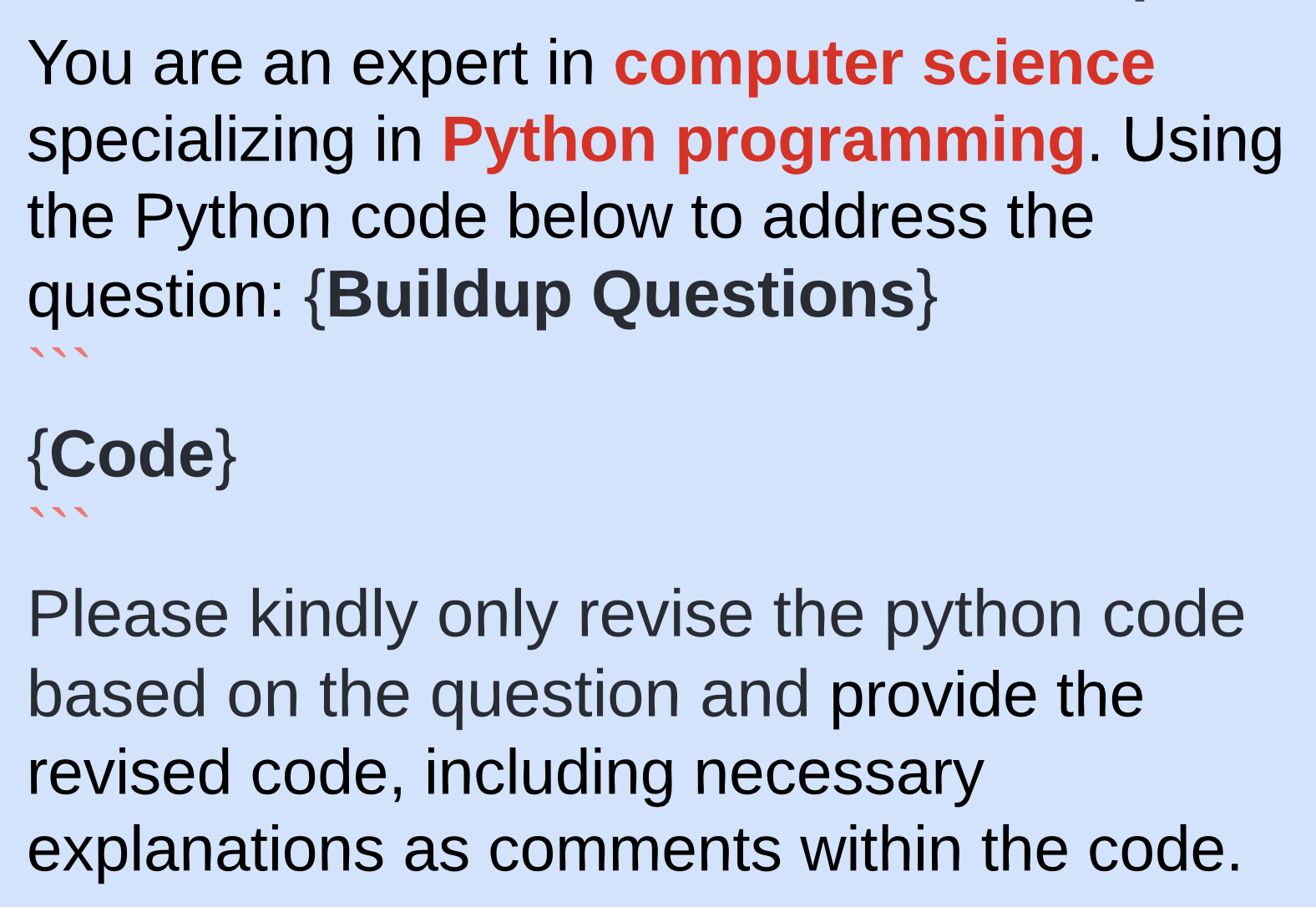}
\caption{Runtime error revision prompts}
\label{code_err_prompt}
\end{figure}
Passing the lint check does not guarantee that the code is bug-free.  Runtime errors, denoted as $err$, may occur during the execution. Additionally, the user might not be satisfied with the execution results of $c_{curr}$. In such cases, they can make requests, denoted as $req$, to revise $c_{curr}$. When such situations happen, users can enter their questions in intuitive style like ``How to fix $\{err\}$?'' and ``I want to $\{req\}$.'' The buildup prompting system will automatically convert these requests into $q_{builtup}$ and update $c_{curr}$ following Equation \ref{eq:buildup} and Figure \ref{code_err_prompt}. This process can also be performed recursively to achieve bug-free and correct code. Combined with recursive linter check process, EduBot can generate the final version of code $c_{final}$ such that $a_i(q_{i})=(c_{final}, r)=a^*_i(q_{i})$.  
\subsubsection{Human-in-the-Loop Inference Chaining}
When EduBot solve the sub-task $q_i$ with either concept questions workflow and coding questions workflow, users can enter their personalized request $q_{i+1}$. $q_{i+1}$ can either be a concept question or a coding question that users believe will be the best approach to solve the complex scenario $Q$. If $q_{i+1}$ is coding questions for adding new features or content based on final code $c_{final}$ of $q_{i}$, $q_{i+1}$ and $c_{final}$ can be directly passed to the buildup prompting system, denoted as $Buildup(q_{i+1},  c_{final})$.

By sequentially conquering all personalized sub-tasks $q$, EduBot can reach the optimal answer $A^*(Q)$. 

\begin{figure}[htbp]
\includegraphics[width=0.45\textwidth]{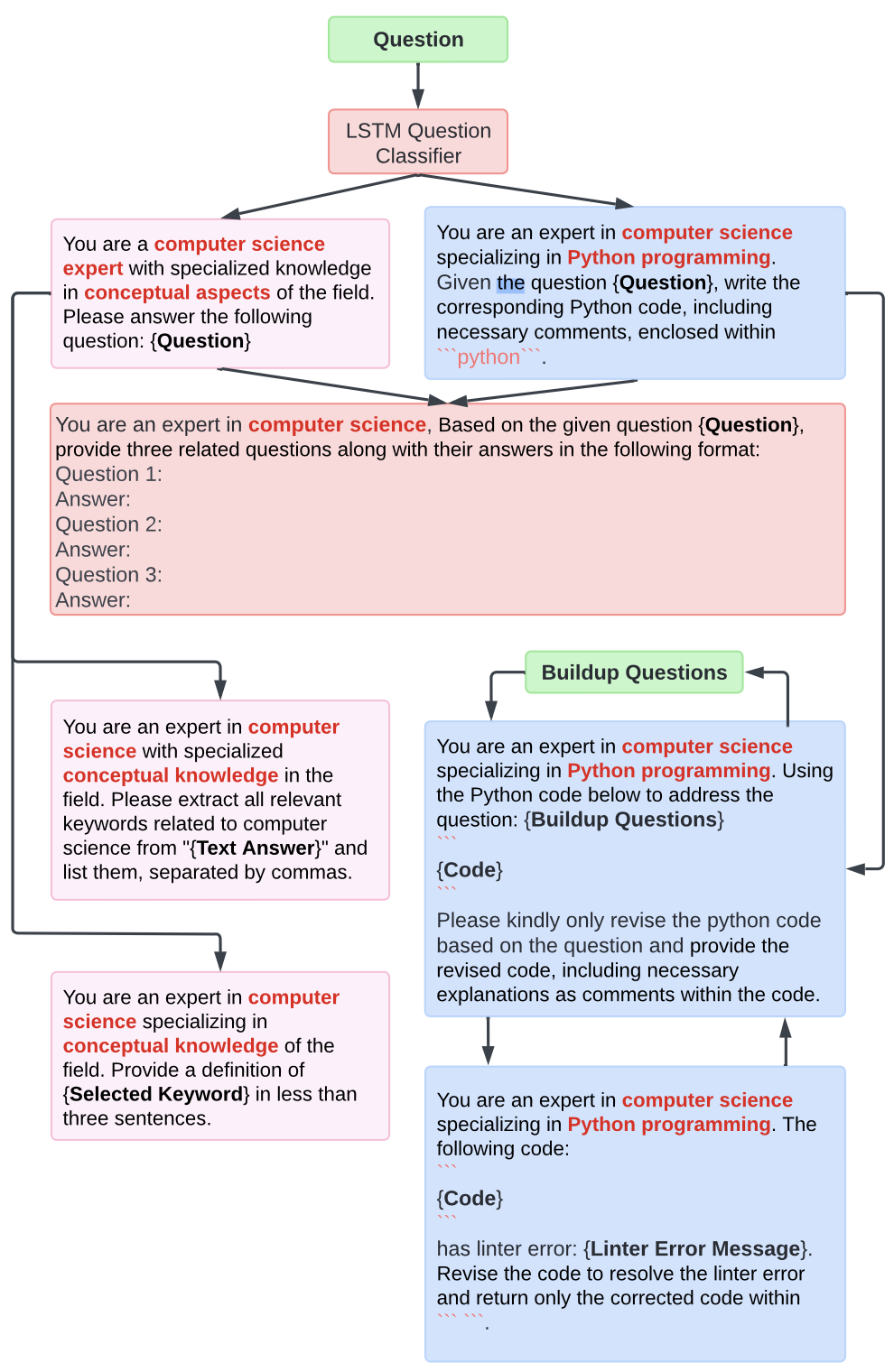}
\caption{Prompting System of EduBot}
\label{prompt_worlflow}
\end{figure}
\subsection{Details of Recursive Prompting System}

EduBot, empowered by designed prompting system shown in Figure \ref{prompt_worlflow}, handles complex conceptual and coding tasks using prompt engineering techniques \cite{chen2023unleashing}. These techniques enable EduBot to achieve robust output and enable recursive interaction without finetuning weights of LLMs. In this section, we will clarify how it works. 

In EduBot, various prompting techniques \cite{liu2023prompting} are used to enhance the quality of responses. Generated Knowledge Prompting \cite{liu2021generated} improves both conceptual and coding questions by iteratively building on previous content and refining solutions. Additionally, prompts inspired by Zero-Shot Chain of Thought \cite {kojima2022large, steinert2023harnessing} regulate output formats such as generating related questions in a specified structure.

Role Prompting \cite{ma2023scope} allows EduBot to adopt specific roles, such as acting as a computer science expert, based on user input. Furthermore, the use of delimiters \cite{OpenAI2024}, such as triple quotes or XML tags, directs the model's focus to specific sections of code, treating other parts as requirements for revision.

\subsection{Machine Learning Methodology}

In EduBot, we use a PyTorch-based LSTM classifier to differentiate between conceptual and coding questions. A dataset of 3,654 distinct questions \cite{LinkAnJarad2022,MujtabaMateen2023} is tokenized and converted into word embeddings. These embeddings are processed through a multi-layered LSTM to extract meaningful features. The extracted features are then passed through a two-layer linear classifier, which outputs a one-hot label to classify the questions as either conceptual (0) or coding-related (1).

We use cross-entropy loss to train the LSTM question classifier which can be expressed as: 
\begin{equation}
L = \sum_{j=1}^{N} t_j \log(p_j) + (1 - t_j)\log(1 - p_j)
\end{equation}
where \( t_i \) is the truth value taking a value 0 or 1 and \( p_i \) is the Softmax probability for the \( i^{th} \) data point of $N$ samples in total.

\section{Experiment Setup}
To evaluate EduBot's ability to solve complex problems with recursive queries, we designed a benchmark suite consisting of twenty working scenarios, each containing multiple sub-tasks.

The benchmark suite is guided by the following principles:
\begin{itemize}
    \item \textbf{Conceptual Understanding}: Each scenario begins with an exploration of fundamental concepts. Understanding the ``why'' behind each topic ensures that learners build a strong theoretical foundation.
    \item \textbf{Algorithmic Implementation}: Learners are tasked with implementing core algorithms or systems in Python. This hands-on approach reinforces learning and provides practical experience.
    \item \textbf{Application to Real-World Problems}: Each scenario includes tasks that apply the learned concepts to real-world situations, emphasizing the practical utility of the algorithms.
    \item \textbf{Incremental Complexity}: Scenarios are designed to progress from understanding basic concepts to implementing complex systems, ensuring a gradual learning curve.
\end{itemize}

Each scenario follows a logical sequence of tasks to ensure comprehensive learning. The general structure is outlined below:
\begin{enumerate}
    \item \textbf{Scenario Introduction}: Provides context and explains the importance of the topic in computer science or data science.
    \item \textbf{Goal Definition}: Breaks down the learning objectives into specific, actionable sub-tasks.
    \item \textbf{Implementation and Practice}: Encourages hands-on coding to apply the concepts in Python.
    \item \textbf{Application and Evaluation}: Applies the implemented solutions to solve problems and evaluates their effectiveness.
\end{enumerate}

The benchmark suite covers various areas of computer science, including fundamental algorithms, machine learning, and real-world problems, to simulate the challenges and questions that learners typically encounter. Each working scenario comprises multiple sub-tasks to complete. A summary of the benchmark suite's key statistics is presented in Table \ref{tab:bench_mark_meta_data}, with further details for each scenario provided in the Appendix.

\begin{table}[h]
  \centering
  \begin{tabular}{lccc}
    \hline
    \textbf{} & \textbf{Algorithm} & \textbf{ML} & \textbf{Real World} \\
    \hline
    Number of Scenarios & 8 & 6 & 6 \\
    Number of Subtasks & 31 & 26 & 22 \\
    Subtasks per Scenario & 3.88 & 4.33 & 3.67 \\
    \hline
  \end{tabular}
  \caption{Statistics of benchmark suites}
  \label{tab:bench_mark_meta_data}
\end{table}

We then pose questions to EduBot based on the scenario and evaluate whether it can successfully accomplish all the sub-goals.

We also record the time spent working with EduBot to accomplish all the sub-goals $q_i$ of the scenario $Q$. If EduBot and the user remain stuck on a sub-task $q_i$ for more than one hour, we consider solving $q_i$ as a failure. EduBot's performance is evaluated based on the ratio of completed sub-tasks $q$ and the time taken to complete each sub-task $q$ within each scenario $Q$. Our experiments are conducted using a range of LLMs, including GPT-3.5-Turbo, GPT-4, and GPT-4-o1.

\section{Results}
In this section, we present the results of our evaluation of EduBot using three different LLM backbones: GPT-3.5, GPT-4, and GPT-4-o1. Our analysis focuses on EduBot's performance across three key metrics: task completion rates, computational efficiency, and the average time required per scenario and per sub-task.

The results reveal trends in EduBot's performance, emphasizing the influence of the underlying LLM's capabilities on its overall effectiveness and efficiency. EduBot demonstrates robust performance in tasks that involve both knowledge acquisition and programming, regardless of the LLM used. Below, we provide a detailed breakdown of these findings.
\subsection{Sub-goals of Benchmark Scenarios}
\begin{figure}[htbp]
\includegraphics[width=0.45\textwidth]{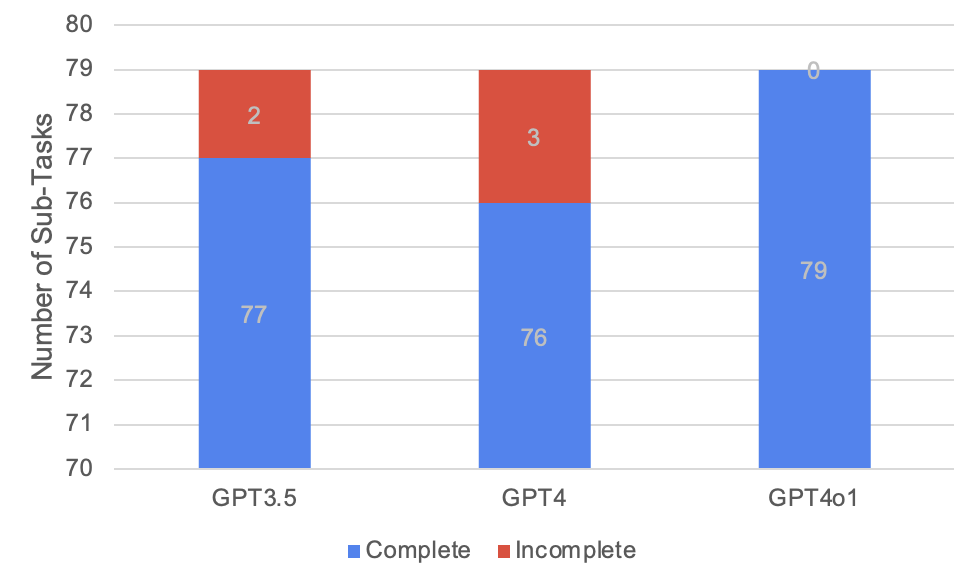}
\caption{Sub-goals of benchmark scenarios in each category completed by EduBot powered by Different LLMs}
\label{subgoals_completness}
\end{figure}
Figure \ref{subgoals_completness} summarizes the number of complete and incomplete sub-tasks by EduBot when powered by different LLMs. EduBot powered by GPT-4-o1 demonstrates the highest effectiveness, completing all 79 sub-tasks with no incompletions. In comparison, EduBot with GPT-3.5 completed 77 sub-tasks with 2 left incomplete, while GPT-4 completed 76 sub-tasks with 3 left incomplete. These results emphasize EduBot's ability to leverage advanced LLMs to solve complex sub-tasks with better usage of the recursive prompts and debugging mechanisms required for multi-step programming tasks.
\subsection{Average Time per Scenario}
\begin{figure}[htbp]
\includegraphics[width=0.45\textwidth]{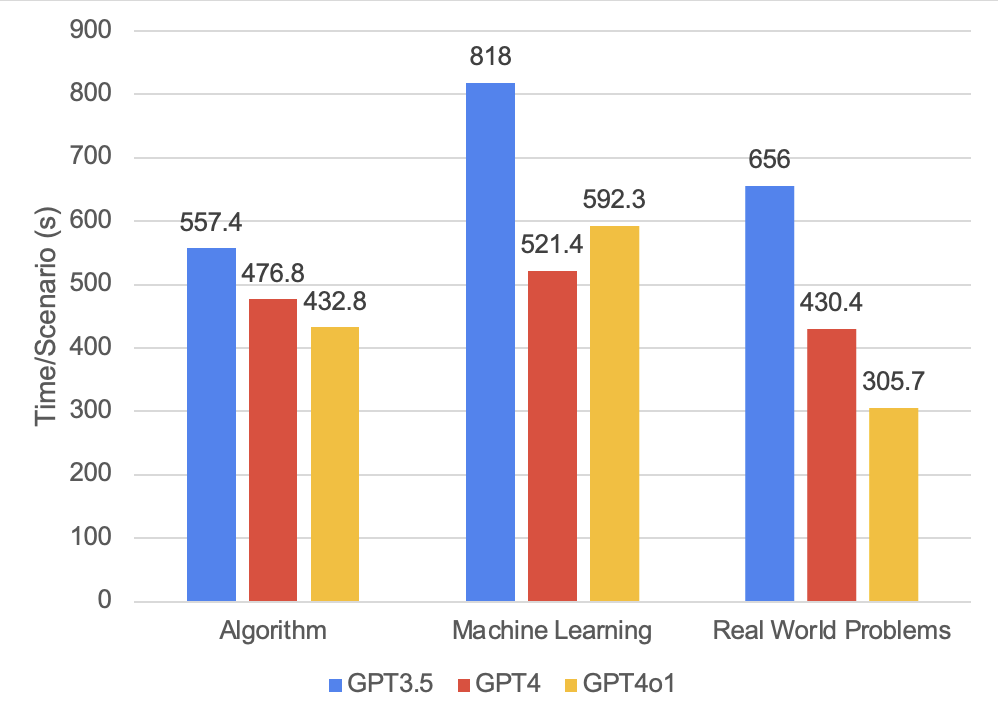}
\caption{Average time needed to complete one benchmark scenario in different categories}
\label{time_per_scen}
\end{figure}
Figure \ref{time_per_scen} compares the average time EduBot requires to complete one benchmark scenario across the three categories: algorithms, machine learning, and real-world problems. GPT-4-o1 consistently outperforms GPT-3.5 and GPT-4 in terms of completion time. For algorithmic scenarios, EduBot with GPT-4-o1 achieves the shortest completion time (377 seconds), followed by GPT-4 (417.3 seconds) and GPT-3.5 (487.8 seconds). In machine learning, GPT-4 achieves the  best performance, taking an average of 521.4 seconds per scenario compared to 818 seconds for GPT-3.5 and 592.3 seconds for GPT-4-o1. However, GPT-4 fails to complete a machine learning scenario involving LSTMs, which GPT-4-o1 successfully completes in approximately 20 minutes. Therefore, while GPT-4 has the lowest average completion time for machine learning scenarios, GPT-4-o1 demonstrates higher task completeness. In the real-world problems category, GPT-4-o1 again outperforms the other LLMs, achieving the best overall performance. These results highlight the efficiency and effectiveness gains achieved by EduBot when powered by more advanced LLMs, particularly GPT-4-o1.
\subsection{Average Time per Sub-goal}
\begin{figure}[htbp]
\includegraphics[width=0.45\textwidth]{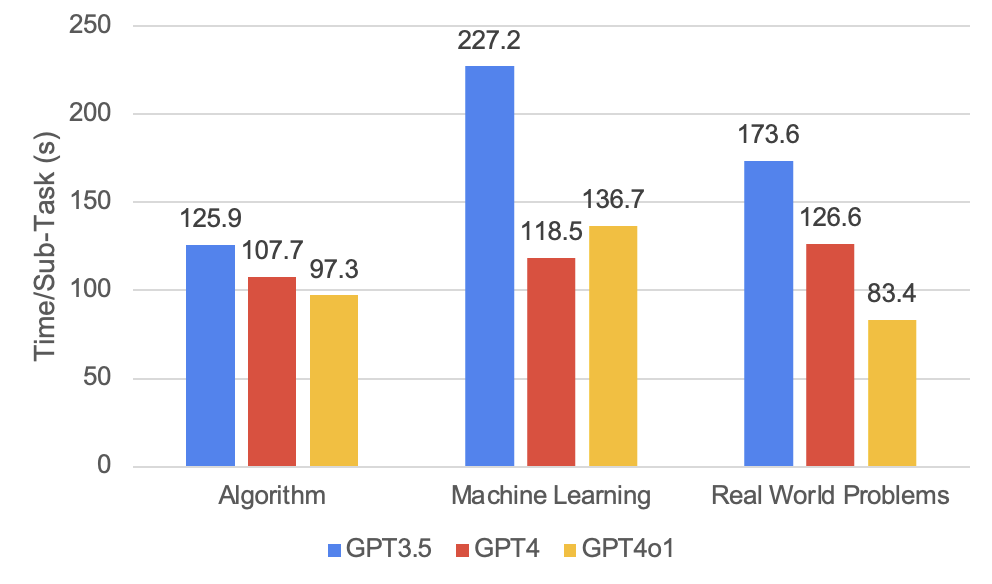}
\caption{Average time needed to complete one sub-goal in different categories}
\label{time_per_goal}
\end{figure}
To better understand EduBot's efficiency, Figure \ref{time_per_goal} illustrates the average time required to complete individual sub-tasks across the three categories. For algorithmic sub-tasks, EduBot with GPT-4-o1 is the most efficient, completing tasks in 97.3 seconds on average, compared to 107.7 seconds for GPT-4 and 125.9 seconds for GPT-3.5. In the machine learning category, GPT-4 gets the best perforamnce in exchange of lower completness of sub-tasks. The efficiency gains are even more pronounced in real-world problem sub-goals, where GPT-4-o1 completes tasks in an average of 83.4 seconds compared to GPT-4’s 126.6 seconds and GPT-3.5’s 173.6 seconds. These findings suggest that GPT-4-o1 enhances EduBot's ability to efficiently solve sub-tasks, particularly in scenarios requiring iterative refinement and debugging.

Based on our experimental results, EduBot demonstrates excellent scalability, leveraging more advanced models to achieve better performance and results.
\section{CONCLUSIONS}
In this paper, we introduced EduBot, an intelligent automated programming assistant powered by Large Language Models (LLMs) to tackle comprehensive programming tasks involving incremental sub-tasks with minimal human intervention. Using a benchmark suite of 20 scenarios spanning algorithms, machine learning, and real-world problems, we demonstrated EduBot's ability to integrate conceptual knowledge learning, recursive prompt-driven programming, and automated debugging into a cohesive workflow. Our results show that EduBot can efficiently complete most scenarios within 20 minutes, with performance and efficiency significantly enhanced when using advanced LLMs like GPT-4-o1. The comparative analysis underscores EduBot's adaptability and robustness, highlighting its potential to extend the utility of pre-trained LLMs for multi-step reasoning, end-to-end code generation, and personalized programming assistance. This work paves the way for further exploration into leveraging LLMs for scalable, intelligent, and recursive coding solutions in educational and real-world applications.

\bibliography{aaai25}
\nocite{cambronero2023flashfill++, ibrahim2023perception, ibrahim2023rethinking}
\appendix
\onecolumn
\section{Scenario Benchmark Suites}\label{sec:benchmark}
\begin{table*}[h]
\centering
\label{table:algo_benchmark}
\begin{tabular}{p{0.15\textwidth}p{0.15\textwidth}p{0.3\textwidth}p{0.3\textwidth}}
\hline
\textbf{Scenario} & \textbf{Type} & \textbf{Scenario Detail} & \textbf{Goal} \\ \hline

Sorting 
& Algorithm  
& I want to learn how various sorting algorithms like Bubble Sort, Merge Sort, and Quick Sort work and understand their time and space complexities. I will practice by implementing these algorithms in Python. 
& 
    1.Understand what sorting is and why it is important in computer science.\newline
    2. Learn about common sorting algorithms like Bubble Sort, Merge Sort, and Quick Sort.\newline
    3. Identify the three most commonly used sorting algorithms and understand their differences.\newline
    4. Implement these sorting algorithms in Python.
\\ \hline

Binary Search Tree (BST) 
& Algorithm 
& I want to understand what a Binary Search Tree is, how it organizes data, and how to traverse it using in-order, pre-order, and post-order methods. I will implement a BST in Python and apply it to solve problems involving search, insertion, and deletion of elements.
& 
    1. Understand what a Binary Search Tree is and how it is structured.\newline
    2. Learn how to write a BST in Python.\newline
    3. Explore different methods for traversing element in a BST.\newline
    4. Implement all traversing methods in Python.
\\ \hline

Dijkstra
& Algorithm 
& I want to study Dijkstra's algorithm to understand how it finds the shortest path in a graph. I will implement it in Python and use it to solve real-world problems like finding the quickest route in a city or optimizing network latency.
& 
    1. Understand the principles behind Dijkstra’s algorithm.\newline
    2. Learn how to implement Dijkstra’s algorithm in Python.\newline
    3. Explore real-world applications of Dijkstra’s algorithm.\newline
    4. Implement a real-world scenario.\newline
\\ \hline

Dynamic Programming
& Algorithm 
& I want to learn the fundamental concepts of dynamic programming, such as breaking problems into sub-problems and using memorization and tabulation. I will practice by solving classic DP problems like Fibonacci, Knapsack, and Longest Common Subsequence in Python.
& 
    1. Understand what Dynamic Programming is and its significance in problem-solving.  \newline
    2. Learn about different approaches to DP, such as top-down and bottom-up.  \newline
    3. Explore situations where bottom-up and top-down approaches are suitable\newline
    4. Write a Python program to find the longest common subsequence of two input strings using DP.\newline
\\ \hline
\end{tabular}
\end{table*}

\begin{table*}[h]
\centering
\label{table:edubot_benchmark}
\label{table:ml_benchmark}
\begin{tabular}{p{0.15\textwidth}p{0.15\textwidth}p{0.3\textwidth}p{0.3\textwidth}}

\hline
\textbf{Scenario} & \textbf{Type} & \textbf{Scenario Detail} & \textbf{Goal} \\ \hline

Heap
& Algorithm 
& I want to understand the structure and functionality of min-heaps and max-heaps. I will implement heaps in Python and solve problems such as priority queue management and heap sort, applying them to real-world scenarios like task scheduling.
& 
    1. Understand the concept of heaps and their properties.\newline
    2. Learn about the real-world applications of heaps, such as priority queues.\newline
    3. Implement a heap from scratch in Python.\newline
    4. Solve a problem to find the top k frequent elements in an array using a heap.
\\ \hline

Linked List
& Algorithm 
& I want to learn about linked lists, including singly and doubly linked lists, and understand their advantages over arrays. I will build linked lists in Python and solve problems such as reversing a linked list or detecting cycles in linked data structures.
& 
    1. Understand what a linked list is and how it differs from arrays.\newline
    2. Learn the differences between singly and doubly linked lists.\newline
    3. Implement a singly linked list from scratch in Python.\newline
    4. Write a program to detect cycles in a linked list.
\\ \hline

Stack
& Algorithm 
& I want to grasp the concept of stacks and their Last-In-First-Out (LIFO) behavior. I will implement stacks in Python and use them to solve problems like checking for balanced parentheses and evaluating postfix expressions.
& 
    1. Understand the stack data structure and its LIFO property.\newline
    2. Implement a stack from scratch in Python.\newline
    3. Solve a problem to check if parentheses in a given string are balanced using a stack.
\\ \hline

Regular Expression
& Algorithm 
& I want to understand the syntax and utility of regular expressions for pattern matching and text processing. I will use Python’s re module to solve problems like extracting data from unstructured text and validating inputs such as email addresses.
& 
    1. Learn what regular expressions are and their importance in pattern matching.\newline
    2. Understand commonly used regular expression patterns.\newline
    3. Explore methods in Python’s `re` module for working with regular expressions.\newline
    4. Write a program to find all 9-digit phone numbers in a given paragraph.
\\ \hline

CSV File + Neural Network
& Machine Learning 
& I want to learn how to work with CSV files in Python by loading and reading them. I will apply machine learning methods like logistic regression to perform binary classification tasks, such as predicting diabetes from medical data.
& 
    1. Understand what a CSV file is and its common uses.\newline
    2. Learn about three effective models for binary classification.\newline
    3. Write Python code to read a CSV file from a given path.\newline
    4. Build a binary classification model to predict diabetes using logistic regression.\newline
    5. Implement a multi-layered neural network in PyTorch for binary classification.
\\ \hline
Regression 
& Machine Learning 
& I want to understand the concept of regression and explore various models, including linear regression and neural networks. I will apply these techniques in Python to predict wine quality and evaluate model performance using metrics like $R^2$ and MAE.
& 
    1. Understand what a regression problem is and its applications.\newline
    2. Learn about different regression models for handling multiple data.\newline
    3. Implement linear regression in Python.\newline
    4. Write a neural network in Python to solve regression problems.
\end{tabular}
\end{table*}

\begin{table*}[h]
\centering
\label{table:edubot_benchmark}
\label{table:ml_benchmark}
\begin{tabular}{p{0.15\textwidth}p{0.15\textwidth}p{0.3\textwidth}p{0.3\textwidth}}

\hline
\textbf{Scenario} & \textbf{Type} & \textbf{Scenario Detail} & \textbf{Goal} \\ \hline
Unsupervised Learning 
& Machine Learning 
& I want to learn about unsupervised learning techniques such as K-Means, Hierarchical Clustering, and DBSCAN. I will apply these methods to classify different types of wines and use dimensionality reduction techniques like PCA to improve clustering accuracy.
& 
    1. Understand the concept of unsupervised learning and its significance.\newline
    2. Learn effective techniques for unsupervised classification.\newline
    3. Implement K-Means clustering in Python.\newline
    4. Solve a wine classification problem using t-SNE and K-Means.\newline
    5. Use PCA with K-Means for dimensionality reduction and classification.
\\ \hline

LSTM
& Machine Learning 
& I want to understand the concept of Long Short-Term Memory (LSTM) networks and how they handle sequence data. I will implement an LSTM model in Python for a simple binary classification task, such as classifying question types.
& 
    1. Learn about neural network models for NLP tasks.\newline
    2. Understand how LSTMs work and why they are suitable for sequential data.\newline
    3. Implement an LSTM model in PyTorch.\newline
    4. Apply the LSTM model to classify conceptual and coding questions into two classes.
\\ \hline

Support Vector Machine
& Machine Learning 
& I want to learn about Support Vector Machines, understand when to use them, and explore their kernel tricks. I will implement SVMs in Python to solve classification problems such as diabetes classification.
& 
    1. Understand what SVM is and its significance in classification.\newline
    2. Learn about common use cases of SVM.\newline
    3. Implement SVM from scratch in Python.\newline
    4. Solve a binary classification diabetes using SVM with sklearn.
\\ \hline

AdaBoost
& Machine Learning 
& I want to understand how the AdaBoost algorithm works for classification tasks. I will implement AdaBoost in Python and apply it to solve problems like predicting diabetes, comparing its performance with other ensemble methods.
& 
    1. Understand what AdaBoost is and how it works.\newline
    2. Learn about the common use cases of AdaBoost in classification.\newline
    3. Implement AdaBoost from scratch in Python.\newline
    4. Apply AdaBoost to predict diabetes in a binary classification task.
\\ \hline

Bank account system (LFU) 
& Real World Problem 
& I want to learn about what is LFU cache and how to implement it and combine with a real world situation. 
& 
    1. Learn what an LFU Cache is and how it works.\newline
    2. Implement an LFU data structure in Python with test cases.\newline
    3. Use LFU to optimize a bank account system in Python.
\\ \hline

Sudoku Solver
& Real World Problem 
& I want to learn the rules of Sudoku and practice solving puzzles manually. Then, I will write a Python program to generate Sudoku puzzles and another to solve them using backtracking or other algorithms.
& 
    1. Understand the rules of Sudoku and how it is played.\newline
    2. Implement a Sudoku puzzle generator in Python.\newline
    3. Learn about the main approaches to solving Sudoku puzzles.\newline
    4. Implement a Sudoku solver using a backtracking algorithm.\newline
    5. Write test code to verify the correctness of a given Sudoku solution.
\\ \hline
\end{tabular}
\end{table*}

\begin{table*}[h]
\centering
\label{table:edubot_benchmark}
\label{table:ml_benchmark}
\begin{tabular}{p{0.15\textwidth}p{0.15\textwidth}p{0.3\textwidth}p{0.3\textwidth}}

\hline
\textbf{Scenario} & \textbf{Type} & \textbf{Scenario Detail} & \textbf{Goal} \\ \hline

Web Scraping 
& Real World Problem 
& I want to learn how to retrieve useful data from a webpage and use this knowledge to find dates for me in the website. 
& 
    1. Understand what web scraping is and why it is useful.\newline
    2. Learn about Python libraries for web scraping, such as Beautiful Soup.\newline
    3. Write a web scraper to extract data from a calendar website.\newline
    4. Retrieve all important dates from the website (https://aideadlin.es/).
\\ \hline

Trie Tree
& Real World Problem 
& I want to understand how Trie Trees work for efficient storage and retrieval of strings. I will implement a Trie in Python to store and manage a dictionary of words, using it to build features like autocomplete.
& 
    1. Learn about data structures for efficiently storing and retrieving strings.\newline
    2. Understand the structure and usage of Trie Trees.\newline
    3. Implement a Trie Tree from scratch in Python.\newline
    4. Write test cases to verify the correctness of the Trie Tree implementation.
\\ \hline

Scientific Calculator
& Real World Problem 
& I want to replace my physical scientific calculator by building a Python program that can perform basic to advanced calculations. I will develop this calculator step by step, eventually handling complex equations and graph plotting.
& 
    1. Understand the basic operations required for a scientific calculator.\newline
    2. Implement a basic calculator in Python supporting addition, subtraction, multiplication, and division.\newline
    3. Extend the calculator to support parentheses.
\\ \hline

Huffman Encoding
& Real World Problem 
& I want to learn about Huffman Encoding, a lossless data compression algorithm. I will study the principles behind Huffman trees and practice building them from character frequency tables. Once I understand the theory, I will implement Huffman Encoding and Decoding in Python. Additionally, I will explore its real-world applications, such as compressing text files.
& 
    1. Understand what Huffman Encoding is and how it is used for lossless data compression.\newline
    2. Learn how to construct a Huffman Tree\newline
    3. Implement Huffman Encoding and Decoding in Python.
\\ \hline

\end{tabular}
\end{table*}
\end{document}